\documentclass[12pt]{article}
\input{tcilatex}
\title{On fermionic tilde conjugation rules and thermal bosonization. Hot and cold
thermofields}
\author{S.E. Korenblit, V.V. Semenov}
\begin{document}
\maketitle
\begin{abstract}
A generalization of Ojima tilde conjugation rules is suggested, which reveals the 
coherent state properties of thermal vacuum state and is useful for the thermofield 
bosonization. 
The notion of hot and cold thermofields is introduced to distinguish different 
thermofield representations giving the correct normal form of thermofield solution for 
finite temperature Thirring model with correct renormalization and anticommutation 
properties. 
\end{abstract}
%%%%%%%%%%%%%%%%%%%%%%%%%%%%%%%%%%%%%%%%%%%%%%%%%%%%%%%%%%%%%%%%%%%%%%%%%%%%%%%%%%%%
%%%%%%%%%%%%%%%%%%%%%%%%%%%%%%%%%%%%%%%%%%%%%%%%%%%%%%%%%%%%%%%%%%%%%%%%%%%%%%%%%%%%%%
\section{Thermodynamics of ideal 1D gases}

From the standard course \cite{isih} it may be easily shown, that equilibrium 
thermodynamics of the free massless bosons in the 1- dimensional box of length $L$ 
coincides with that of the free 
massless spin $1/2$ fermions at the same temperature ${\rm k}_B T=1/\varsigma$ 
only for both zero chemical potentials, $\mu_{(B)}=\mu_{(F)}=0$, 
giving a simplest example of thermal bosonization \cite{gom_ste}, for pressure $P$, 
densities of internal energy ${\cal U}$ and entropy $S$ (with  
$h=2\pi\hbar$, $c$ - speed of light): 
\begin{eqnarray}
&&\!\!\!\!\!\!\!\!\!\!\!\!\!\!\!\!\!\!\!\!
P_{(B),(F)}=\frac{{\cal U}_{(B),(F)}}L=\frac{\pi^2}{3\varsigma^2 hc}, \quad 
\frac{S_{(B),(F)}}{{\rm k}_B L}=\frac{2\pi^2}{3\varsigma hc},\;\;\mbox{ however,}
\label{PBFS} \\
&&\!\!\!\!\!\!\!\!\!\!\!\!\!\!\!\!\!\!\!\!
\mbox{for given densities: }\;\overline{n}_{(B)}=\frac{N_{(B)}}L, \quad 
\overline{n}^\pm_{(F)}=\frac{N^\pm_{(F)}}L\;:
\label{nn} \\
&&\!\!\!\!\!\!\!\!\!\!\!\!\!\!\!\!\!\!\!\!
\mu_{(B)}=\frac 1\varsigma \ln\left(1-e^{-\overline{n}_{(B)}\varsigma hc/2}\right), \quad
\mu^\pm=\frac 1\varsigma \ln\left(e^{\overline{n}^\pm_{(F)}\varsigma hc/2}-1\right).
\label{MuB_MuF}
\end{eqnarray}
This qualitative ``equilibrium'' picture means that both systems for the same 
$\varsigma, L$ have the same $P,{\cal U},S$ and also another thermodynamic potentials. 
The condition $\mu_{(B)}=0$ for arbitrary temperature implies an infinite boson density,  
$\overline{n}_{(B)}\mapsto\infty$, corresponding to specific case of thermodynamic  
limit $N_{(B)}\to\infty$, $L\to\infty$ for bosonic ``picture''. 
The ``equilibrium'' fermion pressure (\ref{PBFS}) actually is a sum of partial ones of 
$N^{+}_{(F)}$ fermions and $N^{-}_{(F)}$ antifermions, with opposite values of chemical 
potentials $\mu^{\pm}=\pm\mu_{(F)}$ and with the charge density 
${\rm Q}_{(F)}/L$ \cite{alv-gom}:
\begin{eqnarray}
&&\!\!\!\!\!\!\!\!\!\!\!\!\!\!\!\!\!\!\!\!  
P_{(F)}=P^+_{(F)} +P^-_{(F)}=
\frac{\pi^2}{3\varsigma^2 hc}+\frac{\mu^2_{(F)}}{hc}, \quad 
\frac {{\rm Q}_{(F)}}{L}=\overline{n}^+_{(F)}-\overline{n}^-_{(F)}=\frac{2\mu_{(F)}}{hc}.
\label{P_xi_P} \\
&&\!\!\!\!\!\!\!\!\!\!\!\!\!\!\!\!\!\!\!\!  
\mbox {So, for any value of $\mu_{(F)},\; \mu_{(B)}$, the ``equilibrium'' Gibbs 
potentials read:}
\nonumber \\
&&\!\!\!\!\!\!\!\!\!\!\!\!\!\!\!\!\!\!\!\!
{\cal G}_{(F)}=N^+_{(F)}\mu^+ +N^-_{(F)}\mu^-=
{\rm Q}_{(F)}\mu_{(F)}=\frac{2L \mu^2_{(F)}}{hc}, \quad 
{\cal G}_{(B)}=N_{(B)}\mu_{(B)}.
\label{Nmu_Nmu} \\
&&\!\!\!\!\!\!\!\!\!\!\!\!\!\!\!\!\!\!\!\!
\mbox{Thus: }\;
{\cal G}_{(F)}\Longrightarrow {\cal G}_{(B)}=0,\;\mbox { only if: }\;
\mu_{(F)}\Rightarrow 0,\;\mbox { with: }\;\overline{n}^+_{(F)}=\overline{n}^-_{(F)}.
\label{N+N-}
\end{eqnarray}
Nevertheless, $\mu_{(F)}=0$, for $\overline{n}^0_{(F)}=2\ln 2/(\varsigma hc)$. 
We want to point out that for nonzero temperature the usual infrared regularization 
parameter $L$ acquires a physical meaning as a macroscopic thermodynamic parameter 
(\ref{PBFS}) of the real or effective ``box size'' of the thermodynamic system 
under consideration. So, the corresponding dependence requires additional care. 

\section{On fermionic tilde conjugation rules}

Following to Ojima \cite{ojima} let us start with simplest fermionic oscillator 
(for one fixed mode $k^1$), which has only two normalized states  $|0\rangle$ and   
$|1\rangle$, with energy $0$ and $\omega$, annihilated/created by fermionic operators 
$b,b^\dagger$: $b|0\rangle=0$, $|1\rangle=b^\dagger|0\rangle$, $\{b,b^\dagger\}=1$, 
$\{b,b\}=0$. The thermal vacuum appears as a normalized sum of tensor products of two 
independent copies of these states:  
$|0\widetilde{0}\rangle=|0\rangle\otimes|\widetilde{0}\rangle$, 
$|1\widetilde{1}\rangle=|1\rangle\otimes|\widetilde{1}\rangle$, weighted with 
corresponding Gibbs and relative phase exponential factors \cite{ojima}, so that for 
$\{b,\widetilde{b}^\#\}=0$, $(\widetilde{b}^\#=\widetilde{b},\widetilde{b}^\dagger)$: 
\begin{eqnarray}
&&\!\!\!\!\!\!\!\!\!\!\!\!\!\!\!\!\!\!\!\!
|0(\varsigma)\rangle_{(F)}=
\frac{|0\widetilde{0}\rangle+e^{i\Phi}e^{-\varsigma\omega/2}|1\widetilde{1}\rangle}
{\left[\langle 0\widetilde{0}|0\widetilde{0}\rangle+e^{-\varsigma\omega}
\langle 1\widetilde{1}|1\widetilde{1}\rangle\right]^{1/2}}\equiv 
\cos\vartheta\left(1+e^{i\Phi}\tan\vartheta\,b^\dagger \widetilde{b}^\dagger \right)
|0\widetilde{0}\rangle ,
\label{OJ_} \\
&&\!\!\!\!\!\!\!\!\!\!\!\!\!\!\!\!\!\!\!\!
|0(\varsigma)\rangle_{(F)}={\cal V}^{-1}_{\vartheta(F)}|0\widetilde{0}\rangle,\; 
\mbox{ where, for: }\;
\tan^2\vartheta(k^1,\varsigma)=e^{-\varsigma\omega}, \quad \omega=\omega(k^1):  
\label{OJ_01} \\
&&\!\!\!\!\!\!\!\!\!\!\!\!\!\!\!\!\!\!\!\!
{\cal V}^{-1}_{\vartheta(F)}=\exp\left\{e^{i\Phi}\tan\vartheta\,G_+\right\}
\exp\left\{-\ln(\cos^2\vartheta)\,G_3\right\}
\exp\left\{-e^{-i\Phi}\tan\vartheta\,G_-\right\},
\label{OJ_1} \\
&&\!\!\!\!\!\!\!\!\!\!\!\!\!\!\!\!\!\!\!\!
\mbox{with: }\;G_+=b^\dagger \widetilde{b}^\dagger, \quad  
G_-=\widetilde{b}b=\left(G_+\right)^\dagger,\quad 
G_3\!=\frac 12\left(b^\dagger b-\widetilde{b}\widetilde{b}^\dagger\right),
\label{OJ_2} \\
&&\!\!\!\!\!\!\!\!\!\!\!\!\!\!\!\!\!\!\!\!
\left[G_+,G_-\right]=2G_3,\quad \left[G_3,G_\pm\right]=\pm G_\pm,\;\mbox{ with: }\;
G_\pm=G_1\pm i G_2, 
\label{OJ_02} \\
&&\!\!\!\!\!\!\!\!\!\!\!\!\!\!\!\!\!\!\!\!
\mbox{ thus: }\;
{\cal V}^{-1}_{\vartheta(F)}=
\exp\left\{\vartheta\left[e^{i\Phi}G_+-e^{-i\Phi}G_-\right]\right\}=
{\cal V}_{-\vartheta(F)}={\cal V}^{\dagger}_{\vartheta(F)}, 
\label{OJ_3}  
\end{eqnarray}
-- is a standard form of operator of the coherent state for group $SU(2)$ 
\cite{perelom}. 
This observation allows to identify the algebra (\ref{OJ_02}) as ``quasispin'' 
algebra \cite{lipkin}, with  the cold vacuum $|0\widetilde{0}\rangle$ as its 
lowest state for representation with ``quasispin'' 1/2, and the state 
$|1\widetilde{1}\rangle$ as the highest one: 
\begin{eqnarray}
&&\!\!\!\!\!\!\!\!\!\!\!\!\!\!\!\!\!\!\!\!
|0\widetilde{0}\rangle \Rightarrow \left|\frac 12,- \frac 12\right\rangle, \quad 
|1\widetilde{1}\rangle \Rightarrow \left|\frac 12, \frac 12\right\rangle,
\label{OJ_4} \\
&&\!\!\!\!\!\!\!\!\!\!\!\!\!\!\!\!\!\!\!\!
G_3\left|\frac 12, \pm \frac 12\right\rangle= \pm \frac 12 
\left|\frac 12, \pm \frac 12\right\rangle, \quad 
G_\pm\left|\frac 12, \pm \frac 12\right\rangle=0. 
\label{OJ_04} 
\end{eqnarray}
The unique arisen arbitrary relative phase $\Phi$ reflects now the fact that the 
quantum state is not the vector, rather the ray. Thus, the thermal vacuum 
(\ref{OJ_01}), as a coherent state \cite{perelom}, is annihilated by operator 
${\cal V}^{-1}_{\vartheta(F)}G_-{\cal V}_{\vartheta(F)}=
\cos^2\vartheta G_- +e^{i\Phi}\sin 2\vartheta G_3-e^{2i\Phi}\sin^2\vartheta G_+=
\,\stackunder{\sim}{b}(\varsigma)b(\varsigma)$, as well as by operators: 
\begin{eqnarray}
&&\!\!\!\!\!\!\!\!\!\!\!\!\!\!\!\!\!\!\!\!
\begin{array}{c}
b(\varsigma)={\cal V}^{-1}_{\vartheta(F)}\, b\, {\cal V}_{\vartheta(F)}=
b\cos\vartheta-\widetilde{b}^\dagger e^{i\Phi}\sin\vartheta, \\
\stackunder{\sim}{b}(\varsigma)=
{\cal V}^{-1}_{\vartheta(F)}\,\widetilde{b}\,{\cal V}_{\vartheta(F)}=
\widetilde{b}\cos\vartheta+b^\dagger e^{i\Phi}\sin\vartheta.
\end{array}
\label{OJ_5} 
\end{eqnarray}
Up to now $\widetilde{b}^\#$ is only notation that does not define any operation. 
To fix it as an operation:  
$\stackunder{\sim}{b}(\varsigma)\mapsto\widetilde{b}(\varsigma)$, one should choose 
the value of $\Phi$. The popular choice $\Phi=0$ leads to complicated tilde conjugation 
rules for the fermionic case, different from the bosonic one \cite{mtu}. The Ojima 
choice $\Phi=-\pi/2$ gives the same rules for both bosonic and fermionic cases 
\cite{ojima}. 
We see now that the choice $\Phi=\pi/2$ is also good and, as well as the original Ojima's 
one, satisfies the properties of 
antilinear homomorphism and the condition 
$\widetilde{\widetilde{b}}(\varsigma)=b(\varsigma)$. 
It seems very convenient for the purposes of bosonization that the tilde operation has 
the same properties for both Fermi and Bose cases. As a byproduct, we observe a useful 
interpretation of the thermal vacuum, defined by Bogoliubov transformation (\ref{OJ_01}), 
as a coherent state, obtained by coherent $SU(2)$ rotation of vacuum states for all Fermi 
oscillators $|0_{k^1}\widetilde{0}_{k^1}\rangle$ as a lowest quasispin states, around 
one and the same unit vector $\vec{\rm u}=(\sin\Phi,\cos\Phi,0)$ onto the different 
angles $=-2\vartheta(k^1)$: ${\cal V}^{-1}_{\vartheta(F)}=
\exp\left[i2\vartheta\left(\vec{\rm u}\cdot\vec{G}\right)\right]$ \cite{perelom}. 

Analogous picture may be obtained for bosonic thermal Bogoliubov transformation 
${\cal V}_{\vartheta(B)}$
leading to connection between the bosonic thermal vacuum and coherent state for the 
discrete series 
representation of group $SU(1,1)$ \cite{perelom}. However, for this case the 
numerator in (\ref{OJ_}) contains a countable number of terms with countable number of 
arbitrary phases $\Phi_n$ \cite{ojima}. The coherent state of the type (\ref{OJ_1}), 
(\ref{OJ_3}) would be obtained only for countable number of coherent choices: 
$\Phi_n\mapsto n\Phi$, 
$n=0,1,2,\ldots$. We did not find a reason to prefer this choice to the usual one 
$\Phi_n=0$ \cite{mtu,ojima}. 

\section{Hot and cold thermofields}

So, at finite temperature, in the framework of thermofield
dynamics \cite{mtu} it is necessary to double the number of
degrees of freedom by providing all the fields $\Psi$ with
their tilde partners $\widetilde{\Psi}$. According to
\cite{mtu}, the resulting theory will be determined by the
Hamiltonian 
$\widehat{H}[\Psi,\widetilde{\Psi}]=H[\Psi]-\widetilde{H}[\widetilde{\Psi}]$,
where $\widetilde{H}[\widetilde{\Psi}] =
H^{*}[\widetilde{\Psi}^{*}]$, with $H[\Psi]=H_{0[\Psi]}(x^0)+H_{I[\Psi]}(x^0)$,  
so that for Thirring model \cite{d_f_z}: 
$\widetilde{H}_{I[\widetilde{\Psi}]}=H_{I[\widetilde{\Psi}]}$,
and $\widetilde{H}_{0[\widetilde{\Psi}]}=-H_{0[\widetilde{\Psi}]}$. 
Though the substitution like (\ref{OJ_5}), for the free massless Dirac 
thermofields, $\chi(x)\mapsto\chi(x,\varsigma)$, also does not change \cite{mtu} the form 
of the free operator: 
$\widehat{H}_{0}[\chi,\widetilde{\chi}]=H_{0}[\chi]-\widetilde{H}_{0}[\widetilde{\chi}]$, 
these free fields, generally speaking, are not now the physical 
fields of this QFT model \cite{blot,alv-gom}, and, as is well known \cite{blot,mtu}, 
each term ${H}[\Psi]$ in $\widehat{H}[\Psi,\widetilde{\Psi}]$ must be equivalent in a 
weak sense to the free Hamiltonian of massless (pseudo) 
scalar fields $(\phi(x))$, $\varphi(x)$, at least, at zero temperature, $T=0$. 

For any functional ${\cal F}\left[\Psi\right]$ of Heisenberg fields (HF) in the given 
representation of physical fields $\psi(x)$, i.e. for dynamical mapping (DM) 
$\Psi(x)=\Upsilon[\psi(x)]$ \cite{mtu} at zero temperature, being interested in the 
matrix elements on the thermal vacuum of the type: 
\begin{eqnarray}
&&\!\!\!\!\!\!\!\!\!\!\!\!\!\!\!\!\!\!\!\!
\langle 0(\varsigma)|{\cal F}\left[\Psi(x)\right]|0(\varsigma)\rangle=
\langle 0\widetilde{0}|{\cal V}_{\vartheta}{\cal F}\left[\Psi(x)\right] 
{\cal V}^{-1}_{\vartheta}|0\widetilde{0}\rangle=
\langle 0\widetilde{0}|{\cal F}\left[{\cal V}_{\vartheta}\Psi(x)
{\cal V}^{-1}_{\vartheta}\right]\!|0\widetilde{0}\rangle, 
\label{VFV_1} \\
&&\!\!\!\!\!\!\!\!\!\!\!\!\!\!\!\!\!\!\!\!
\mbox{we come to formal mapping: }\; 
\nonumber \\
&&\!\!\!\!\!\!\!\!\!\!\!\!\!\!\!\!\!\!\!\!
{\cal V}_{\vartheta}\Psi(x){\cal V}^{-1}_{\vartheta}=\Psi(x,[-]\varsigma)=
\Upsilon\left[{\cal V}_{\vartheta}\psi(x){\cal V}^{-1}_{\vartheta}\right]=
\Upsilon\left[\psi(x,[-]\varsigma)\right],
\label{VPV_1} \\
&&\!\!\!\!\!\!\!\!\!\!\!\!\!\!\!\!\!\!\!\!
\mbox{onto the ``cold'' physical thermofield: }\; 
\psi(x,[-]\varsigma)={\cal V}_{\vartheta}\psi(x){\cal V}^{-1}_{\vartheta},
\label{cold_psi}
\end{eqnarray}
essentially with the same coefficient functions, as for the initial DM 
$\Psi(x)=\Upsilon[\psi(x)]$, that, contrary to \cite{mtu, ojima}, thus transferring 
so all the temperature dependence from the state (\ref{OJ_01}) onto these ``cold'' 
physical thermofields. 
However, to compute the matrix element (\ref{VFV_1}) it is necessary to substitute into 
the r.h.s. of (\ref{VFV_1}), (\ref{VPV_1}) the cold physical thermofields 
(\ref{cold_psi}) again in terms of the initial physical fields $\psi(x)$ via obtained 
from (\ref{cold_psi}) their linear combinations, analogous (but not the same!) to Eqs. 
(\ref{OJ_5}), and reorder again the so obtained operator with respect 
to the initial physical fields $\psi(x)$. The same operations also convert the formal 
mapping (\ref{VPV_1}) into temperature dependent DM over the cold vacuum 
$|0\widetilde{0}\rangle$, and precisely in such sense we call further the r.h.s. of 
(\ref{VPV_1}) again as a new DM $\widehat{\Upsilon}$: 
$\Psi(x,[-]\varsigma)=\Upsilon\left[\psi(x,[-]\varsigma)\right]=
\Upsilon\left[{\cal V}_{\vartheta}\psi(x){\cal V}^{-1}_{\vartheta}\right]
\Rightarrow \widehat{\Upsilon}\left[[-]\varsigma; \psi(x)\right]=
\widehat{\Upsilon}\left[[-]\varsigma; c(k^1),\widetilde{c}(k^1)\right]$, or   
e.g. $c(k^1)\mapsto b_{k^1}$. 

On the contrary, the standard computation way \cite{mtu, ojima} implies the substitution 
of the inverse to (\ref{OJ_5}) linear expressions of physical fields 
$\psi(x)={\cal V}_{\vartheta}\psi(x,[+]\varsigma){\cal V}^{-1}_{\vartheta}$ 
in terms of the ``hot'' physical thermofields,  
$\psi(x,[+]\varsigma)={\cal V}^{-1}_{\vartheta}\psi(x){\cal V}_{\vartheta}$, given by 
(\ref{OJ_5}), into the l.h.s. of (\ref{VFV_1}) and reordering the so obtained operator 
with respect to {\bf this}  
hot physical thermofield over the thermal (``hot'') vacuum (\ref{OJ_01}). 
Of course, such operations give the new DM $\widehat{\Upsilon}$ for the initial HF over 
this hot -- thermal vacuum \cite{mtu}: 
$\Psi(x)=\Upsilon[\psi(x)]=
\Upsilon[{\cal V}_{\vartheta}\psi(x,[+]\varsigma){\cal V}^{-1}_{\vartheta}]
\Rightarrow \widehat{\Upsilon}\left[[+]\varsigma;\psi(x,[+]\varsigma)\right]=
\widehat{\Upsilon}\left[[+]\varsigma; c(k^1,[+]\varsigma),
\widetilde{c}(k^1,[+]\varsigma)\right]$. 
We want to point out that this field does not equal to 
$\Psi(x,[+]\varsigma)={\cal V}^{-1}_{\vartheta}\Psi(x){\cal V}_{\vartheta}$, which 
will appear below as a byproduct of our further consideration.
To avoid some ambiguities \cite{abr,abr3} one should carefully distinguish the hot 
and cold physical thermofields $\psi(x,[\pm]\varsigma)$ over corresponding vacua.  

The kinematic independence of tilde-conjugate fields $\widetilde{\Psi}$ means: 
\begin{eqnarray}
\left\{\Psi_\xi(x),\widetilde{\Psi}^{\#}_{\xi'}(y)\right\}\Bigr|_{x^0= y^0}=0,\qquad 
\left\{\Psi_\xi(x),\widetilde{\Psi}^{\#}_{\xi'}(y)\right\}\Bigr|_{(x-y)^2<0}=0,
\label{T_vbnmei4}
\end{eqnarray}
and corresponds to above independence of their Hamiltonians and their HEqs. This allows 
to consider a solution only for the one of them. Since the thermal transformations 
${\cal V}_{\vartheta(F)}$, ${\cal V}_{\vartheta(B)}$ are not depend on coordinates and 
time, they can be applied directly to Eqs. (\ref{T_vbnmei4}) and zero temperature HEq of 
Thirring model \cite{ks_t}, resulting\footnote{Here:
$x^\mu=\left(x^0,x^1\right)$; $x^0=t$; $\hbar=c=1$;
$\partial_\mu=\left(\partial_0,\partial_1\right)$; for
$g^{\mu\nu}$: $g^{00}=-g^{11}=1$; for $\epsilon^{\mu\nu}$:
$\epsilon^{01}=-\epsilon^{10}=1$; 
$\overline{\Psi}(x)=\Psi^\dagger(x)\gamma^0$; 
$\gamma^0=\sigma_1$, $\gamma^1=-i\sigma_2$, 
$\gamma^5=\gamma^0\gamma^1=\sigma_3$, $\gamma^\mu\gamma^5=-\epsilon^{\mu\nu}\gamma_\nu$, 
where $\sigma_i$ -- Pauli matrices, and $I$ -- unit matrix; $x^\xi = x^0 + \xi x^1$, 
$2\partial_\xi =2{\partial}/{\partial x^\xi}=\partial_0+\xi\partial_1$, 
$P^1=-i\partial_1$, $E(P^1)=\gamma^5P^1$; summation over repeated $\xi=\pm$, is nowhere 
implied. The label $[\pm]$ is omitted, where it is not important:  
$J_{(\Psi)}^\nu(x,\varsigma)\mapsto
\overline{\Psi}(x,\varsigma)\gamma^\nu\Psi(x,\varsigma)$.} 
again to the same Eqs. (\ref{T_vbnmei4}) and HEqs for the new HF 
$\Psi(x,[\pm]\varsigma)$ like (\ref{VPV_1}): 
\begin{eqnarray}
&&\!\!\!\!\!\!\!\!\!\!\!\!\!\!\!\!\!\!\!\!
i\partial_0\Psi(x,\varsigma)=\left[\Psi(x,\varsigma),  
\widehat{H}[\Psi,\widetilde{\Psi}]\,\right]=
\left[E(P^1)+g\gamma^0\gamma_\nu J_{(\Psi)}^\nu(x,\varsigma)
\right]\Psi(x,\varsigma),
\label{T_45bn6l4} \\
&&\!\!\!\!\!\!\!\!\!\!\!\!\!\!\!\!\!\!\!\!
2\partial_\xi \Psi_\xi(x,\varsigma)=
-igJ^{-\xi}_{(\Psi)}(x,\varsigma)\Psi_\xi(x,\varsigma), \;\; 
2\partial_\xi\widetilde{\Psi}_\xi(x,\varsigma)=
ig \widetilde{J}^{-\xi}_{(\widetilde{\Psi})}(x,\varsigma)
\widetilde{\Psi}_\xi(x,\varsigma),
\label{T_45bn6l5}
\end{eqnarray}
-- for each $\xi$ - component of the field, that are also formally related to the 
corresponding current components as:
\begin{eqnarray}
J_{(\Psi)}^\xi (x,\varsigma)=J_{(\Psi)}^0 (x,\varsigma)+\xi J_{(\Psi)}^1(x;\varsigma)
\longmapsto 2\Psi_\xi^\dagger(x,\varsigma)\Psi_\xi (x,\varsigma).
\label{T_vbnmei8}
\end{eqnarray}
Thus, to integrate these HEqs we can sequentially repeat all the steps of our previous 
works \cite{ks_t} for $T=0$. Applying the same arguments based on the currents 
conservation: $\partial_\xi J_{(\Psi)}^\xi (x,\varsigma)=0$, $\xi=\pm$, we come to the 
same weak linearization conditions (here ``w'' means weak equality): 
\begin{eqnarray}
&&\!\!\!\!\!\!\!\!\!\!\!\!\!\!\!\!\!\!
\gamma^0\gamma_\nu J_{(\Psi)}^\nu(x,\varsigma)
\stackrel{\rm w}{\longmapsto}
\frac{\beta}{2\sqrt{\pi}}\gamma^0\gamma_\nu
\widehat{J}_{(\chi)}^\nu(x,\varsigma),
\label{T_ns94m61} \\
&&\!\!\!\!\!\!\!\!\!\!\!\!\!\!\!\!\!\!
\widehat{J}_{(\chi)}^\nu(x,\varsigma)=
\lim\limits_{\varepsilon,(\widetilde{\varepsilon})\rightarrow 0}
\widehat{J}_{(\chi)}^\nu\left(x;\varepsilon
(\widetilde{\varepsilon}),\varsigma\right)\,
\equiv\, :J_{(\chi)}^\nu(x,\varsigma):\,,\;\mbox{ with }\;Z_{(\chi)}(a)=1, 
\label{T_ns94m62} 
\end{eqnarray}
that for the same subsequently normal ordered (renormalized) current:
\begin{eqnarray}
&&\!\!\!\!\!\!\!\!\!\!\!\!\!\!\!\!\!\!
J^0_{(\Psi)} (x,\varsigma) \longmapsto
\lim\limits_{\widetilde{\varepsilon} \rightarrow 0}
\widehat{J}^0_{(\Psi)}(x;\widetilde{\varepsilon},\varsigma)
=\widehat{J}^0_{(\Psi)}(x,\varsigma),
\label{T_bos-111}\\
&&\!\!\!\!\!\!\!\!\!\!\!\!\!\!\!\!\!\!
J^1_{(\Psi)} (x,\varsigma) \longmapsto
\lim\limits_{\varepsilon \rightarrow 0}
\widehat{J}^1_{(\Psi)}(x;\varepsilon,\varsigma)=
\widehat{J}^1_{(\Psi)}(x,\varsigma),
\label{T_bos-0111}\\
&&\!\!\!\!\!\!\!\!\!\!\!\!\!\!\!\!\!\!
\mbox{where at first: }\; \widetilde{\varepsilon}^0 =
\varepsilon^1\rightarrow 0,\;
\mbox{ when: }\; \widetilde{\varepsilon}^1 =\varepsilon^0,\;\;
\varepsilon^2=-\widetilde{\varepsilon}^2>0,\;\mbox{ for:}
\label{T_K_E} \\
&&\!\!\!\!\!\!\!\!\!\!\!\!\!\!\!\!\!\!
\widehat{J}^\nu_{(\Psi)}(x;a,\varsigma)\!=\!
Z^{-1}_{(\Psi)}(a)\!\left[\overline{\Psi}(x + a,\varsigma)
\gamma^\nu \Psi (x,\varsigma)\!-\!
\langle 0\widetilde{0}|\overline{\Psi}(x + a,\varsigma)\gamma^\nu 
\Psi(x,\varsigma)|0\widetilde{0}\rangle \!\right]\!, 
\label{T_K_Z} 
\end{eqnarray}
with (the same) appropriate renormalization constant $Z_{(\Psi)}(a)$, 
leads again to the linearization of both equations (\ref{T_45bn6l4}), (\ref{T_45bn6l5}) 
in the representation of the free physical fields $\chi(x,\varsigma)$. So, that again 
the strong operator bosonization rules for the free field only are necessary: 
\begin{eqnarray}
&&\!\!\!\!\!\!\!\!\!\!\!\!\!\!\!\!\!\!
\widehat{J}_{(\chi)}^\mu (x,\varsigma)=\frac{1}{\sqrt{\pi}}
\partial^\mu\varphi(x,\varsigma)=
- \frac{1}{\sqrt{\pi}}\epsilon^{\mu\nu}\partial_\nu \phi
(x,\varsigma), 
\label{T_nweyi019} \\
&&\!\!\!\!\!\!\!\!\!\!\!\!\!\!\!\!\!\! 
\widehat{J}_{(\chi)}^{-\xi} (x,\varsigma) = \frac{2}{\sqrt{\pi}}
\partial_{\xi} \varphi^{\xi}\left(x^{\xi},\varsigma\right)\,.
\label{T_nweyi19} 
\end{eqnarray}
The thermofields $\varphi(x,\varsigma)$ and $\phi(x,\varsigma)$ are 
defined in (\ref{T_ppp}) below as unitarily inequivalent representations of the 
massless scalar and pseudoscalar Klein-Gordon fields: 
$\partial_\mu\partial^\mu\varphi(x,\varsigma)=0$, and 
$\partial_\mu\partial^\mu\phi(x,\varsigma)=0$, and are taken mutually 
dual and coupled by the symmetric integral relations: $\varepsilon(s)={\rm sgn}(s)$, 
\begin{eqnarray}
&&
\left. \begin{array}{c}\phi(x,\varsigma) \\
\varphi(x,\varsigma)\end{array}\right\}
=-\frac{1}{2}\int\limits_{-\infty}^\infty dy^1
\varepsilon \left(x^1-y^1\right)\partial_0
\left\{\begin{array}{c}\varphi\left(y^1,x^0,\varsigma\right), \\
\phi\left(y^1,x^0,\varsigma \right), \end{array}\right.
\label{T_K_5}
\end{eqnarray}
that implies the conditions: 
$\varphi(-\infty,x^0,\varsigma)+\varphi(\infty,x^0,\varsigma)=
\phi(-\infty,x^0,\varsigma)+\phi(\infty,x^0,\varsigma)=0$. 
The corresponding conserved charges read: 
\begin{eqnarray}
&&\!\!\!\!\!\!\!\!\!\!\!\!\!\!\!\!\!\!
\left. \begin{array}{c} O(\varsigma) \\ O_5(\varsigma)\end{array}\right\}=
\lim_{L\to\infty}
\int\limits_{-\infty}^\infty dy^1\Delta\left(\frac{y^1}L\right)\partial_0
\left\{\begin{array}{c}\varphi\left(y^1,x^0,\varsigma\right) \\
\phi\left(y^1,x^0,\varsigma\right) \end{array}\right\}
\label{T_K_O0} \\
&&\!\!\!\!\!\!\!\!\!\!\!\!\!\!\!\!\!\!
\stackunder{\Delta=1}{\Longrightarrow}
\left\{\begin{array}{c}\phi(-\infty, x^0,\varsigma)-\phi(\infty,x^0,\varsigma) \\
\varphi(-\infty,x^0,\varsigma)-\varphi(\infty,x^0,\varsigma),\end{array}\right. 
\label{T_K_O} 
\end{eqnarray}
where $\Delta(y^1/L)$ is the volume cut-off regularization function with the Fourier 
image $\delta_L(k^1)$ (\ref{DDddDD}). Right ($\xi=-$) and left ($\xi=+$) thermofields 
$\varphi^{\xi}\left(x^{\xi},\varsigma\right)$ and their charges $Q^\xi(\varsigma)$ are 
defined by similar to \cite{blot} linear combinations:
\begin{eqnarray}
&&
\varphi^\xi\left(x^\xi,\varsigma\right)=
\frac{1}{2}\left[\varphi(x,\varsigma)-\xi\phi(x,\varsigma)\right], \;
\mbox { for: }\; \xi=\pm ,
\label{T_K_7} \\
&&
Q^\xi (\varsigma)=\frac{1}{2}\left[O(\varsigma)-\xi {O}_5(\varsigma)\right]=
\pm 2\varphi^\xi\left(x^0\pm\infty,\varsigma\right),
\label{T_K_7_Q}
\end{eqnarray}
These fields obey the temperature independent commutation relations:
\begin{eqnarray}
&&\!\!\!\!\!\!\!\!\!\!\!\!\!\!\!\!\!\!
\left[\varphi(x,\varsigma),\partial_0 \varphi (y,\varsigma)\right]
\bigr|_{x^0=y^0}=
\left[\phi(x,\varsigma),\partial_0\phi(y,\varsigma)\right]
\bigr|_{x^0=y^0}=i\delta(x^1-y^1),
\label{T_K_8_1} \\
&&\!\!\!\!\!\!\!\!\!\!\!\!\!\!\!\!\!\!
\left[\varphi(x,\varsigma), \varphi (y,\varsigma)\right]=
\left[\phi(x,\varsigma),\phi(y,\varsigma)\right]=
-i\frac{\varepsilon(x^0-y^0)}2\theta\left((x-y)^2\right),
\label{T_K_8} \\
&&\!\!\!\!\!\!\!\!\!\!\!\!\!\!\!\!\!\!
\left[\varphi^\xi\left(s,\varsigma\right),\varphi^{\xi'}
\left(\tau,\varsigma\right)\right]=
-\frac{i}{4}\varepsilon(s - \tau)\delta_{\xi, \xi'}, \quad
\left[\varphi^\xi(s,\varsigma), Q ^{\xi'}(\varsigma)\right]=
\frac{i}{2}\delta_{\xi,\xi'}.
\label{T_K_9}
\end{eqnarray}
The similar commutation relations take place for their tilde-partner, that 
remain kinematically independent also at finite temperature: 
$[A(\varsigma),\widetilde{B}(\varsigma)]=0$. So, up to now we cannot 
distinguish the hot and cold physical thermofields. 

The kinematic independence of the tilde-partners fails and the difference between 
the hot and cold physical thermofields appears on going to the ``frequency'' parts of 
corresponding fields $\varphi^{\xi(\pm)}\left(x^\xi,\varsigma\right)$, and their charges 
$Q ^{\xi(\pm)}(\varsigma)$. It manifests itself in the commutators of annihilation $(+)$ 
and creation $(-)$ (frequency) parts, defined by annihilation and creation operators 
over the initial cold vacuum $|0\widetilde{0}\rangle$ for the pseudoscalar fields 
\cite{ks_t}: ${\cal P}c(k^1){\cal P}^{-1}=-c(-k^1)$, 
$[c(k^1),c^\dagger(q^1)]=2\pi 2k^0\delta(k^1-q^1)$, $c(k^1)|0\rangle=0$, 
$c(k^1)|0\widetilde{0}\rangle=\widetilde{c}(k^1)|0\widetilde{0}\rangle=0$, 
for both hot $[+]$, and cold $[-]$ thermofields, in the form:
\begin{eqnarray}
&&\!\!\!\!\!\!\!\!\!\!\!\!\!\!\!\!\!\!
|0 (\varsigma)\rangle = {\cal V}_{\vartheta(B)}^{-1}
|0\widetilde{0}\rangle \equiv {\cal V}_{(B)} [-\vartheta]
|0\widetilde{0}\rangle, \;\; \tanh^2\vartheta=e^{-\varsigma k^0},\;\;
 \vartheta=\vartheta(k^1;\varsigma), 
\label{vac_vac} \\
&&\!\!\!\!\!\!\!\!\!\!\!\!\!\!\!\!\!\!
\varphi(x;[\pm]\varsigma)=
{\cal V}_{\vartheta(B)}^{\mp 1} \varphi (x) {\cal V}_{\vartheta(B)}^{\pm 1}
\Longrightarrow \varphi^{(+)} (x; [\pm] \varsigma) +
\varphi^{(-)}(x;[\pm]\varsigma),
\label{T_ppp}
\end{eqnarray}
and so on for all other fields $\phi(x), \varphi^\xi (x^\xi), Q^\xi,...$, with 
corresponding Fourier expansions and commutators. Below we put corresponding $\pm$ into 
respective brackets, and $k^0 = |k^1|$: 
\begin{eqnarray}
&&\!\!\!\!\!\!\!\!\!\!\!\!\!\!\!\!\!\!
\varphi^{\xi(+)} \left(x^\xi; [\pm]\varsigma\right) = -
\frac{\xi}{2\pi} \int\limits_{-\infty}^\infty \frac{d k^1}{2 k^0}
\theta \left(- \xi k^1\right) \left[\cosh \vartheta c
\left(k^1\right) e^{- i k^0 x^\xi}\mp \right.
\label{phi_T_p} \\
&&\!\!\!\!\!\!\!\!\!\!\!\!\!\!\!\!\!\!
\left. \mp \sinh \vartheta
\widetilde{c} \left(k^1\right) e^{i k^0 x^\xi} \right],\quad 
\varphi^{\xi (-)} \left(x^\xi; [\pm]\varsigma\right) =
\left\{\varphi^{\xi (+)} \left(x^\xi;[\pm]
\varsigma\right)\right\}^\dagger, 
\label{phi_T_m} \\ 
&&\!\!\!\!\!\!\!\!\!\!\!\!\!\!\!\!\!\!
\widetilde{\varphi}^{\xi (+)} \left(x^\xi; [\pm]\varsigma\right) = -
\frac{\xi}{2\pi} \int\limits_{-\infty}^\infty \frac{d k^1}{2 k^0}
\theta \left(- \xi k^1\right) \left[\cosh \vartheta \widetilde{c}
\left(k^1\right) e^{i k^0 x^\xi}\mp \right.
\label{phi_Tt_p} \\
&&\!\!\!\!\!\!\!\!\!\!\!\!\!\!\!\!\!\!
\left. \mp \sinh \vartheta
c \left(k^1\right) e^{- i k^0 x^\xi} \right],\quad 
\widetilde{\varphi}^{\xi (-)} \left(x^\xi; [\pm]\varsigma\right) =
\left\{\widetilde{\varphi}^{\xi (+)} \left(x^\xi;[\pm]
\varsigma\right)\right\}^\dagger, 
\label{phi_Tt_m} \\
&&\!\!\!\!\!\!\!\!\!\!\!\!\!\!\!\!\!\!
Q^{\xi (+)} ([\pm]\varsigma) = \lim_{L \rightarrow \infty} i
\frac{\xi}{2} \int\limits_{-\infty}^\infty d k^1
\theta \left(-\xi k^1\right) \left[\cosh \vartheta c
\left(k^1\right) e^{- i k^0 \widehat{x}^0} \pm \right.
\label{QQ_T_p} \\
&&\!\!\!\!\!\!\!\!\!\!\!\!\!\!\!\!\!\!
\left. \pm \sinh \vartheta \widetilde{c}
\left(k^1\right) e^{i k^0 \widehat{x}^0} \right] \delta_L \left(k^1\right),\quad 
Q^{\xi (-)} ([\pm]\varsigma) = \left\{Q^{\xi (+)}
([\pm]\varsigma)\right\}^\dagger.
\label{QQ_T_m} \\
&&\!\!\!\!\!\!\!\!\!\!\!\!\!\!\!\!\!\!
\widetilde{Q}^{\xi (+)} ([\pm]\varsigma) = \lim_{L \rightarrow \infty} - i
\frac{\xi}{2} \int\limits_{-\infty}^\infty d k^1
\theta \left(-\xi k^1\right) \left[\cosh \vartheta \widetilde{c}
\left(k^1\right) e^{i k^0 \widehat{x}^0} \pm \right.
\label{QQ_Tt_p} \\
&&\!\!\!\!\!\!\!\!\!\!\!\!\!\!\!\!\!\!
\left. \pm \sinh \vartheta c
\left(k^1\right) e^{- i k^0 \widehat{x}^0} \right] \delta_L \left(k^1\right),\quad 
\widetilde{Q}^{\xi (-)} ([\pm]\varsigma) = \left\{\widetilde{Q}^{\xi (+)}
([\pm]\varsigma)\right\}^\dagger.
\label{QQ_Tt_m} 
\end{eqnarray}
Here the $\widehat{x}^0$ -- dependence of charge frequency parts is fictitious and 
unphysical. It is the artifact of space regularization (\ref{T_K_O0}) and should be 
eliminated at the end of calculation. Only for hot $[+]$ thermofields one has:
\begin{eqnarray}
&&\!\!\!\!\!\!\!\!\!\!\!\!\!\!\!\!\!\!
\langle 0 (\varsigma)|\varphi^\xi(s;[+]\varsigma)\varphi^{\xi'}(\tau;[+]\varsigma)
|0(\varsigma)\rangle=\langle 0|\varphi^{\xi}(s)\varphi^{\xi'}(\tau)|0\rangle=
\label{hot_DW0} \\
&&\!\!\!\!\!\!\!\!\!\!\!\!\!\!\!\!\!\!
= \left[\varphi^{\xi(+)}(s),\varphi^{\xi'(-)}(\tau)\right]
= \frac{\delta_{\xi,\xi'}}{i} D^{(-)} (s-\tau),
\label{hot_DW} 
\end{eqnarray}
(here $D^{(-)}(s)=\lim_{\varsigma \rightarrow \infty}{\cal D}^{(-)}(s,\varsigma;\mu_1)$) 
but for both of them: 
\begin{eqnarray}
&&\!\!\!\!\!\!\!\!\!\!\!\!\!\!\!\!\!\!
\langle 0\widetilde{0}|\varphi^\xi(s;[\pm]\varsigma)\varphi^{\xi'}(\tau;[\pm]\varsigma)
|0\widetilde{0}\rangle = 
\left[\varphi^{\xi(+)}(s;[\pm]\varsigma),\varphi^{\xi'(-)}(\tau;[\pm]\varsigma)\right],
\label{p_DW} \\
&&\!\!\!\!\!\!\!\!\!\!\!\!\!\!\!\!\!\!
\left[\varphi^{\xi (\pm)} \left(s; [\pm]\varsigma\right), \varphi^{\xi'
(\mp)} \left(\tau; [\pm]\varsigma\right)\right]
= (\pm 1)\frac{\delta_{\xi,\xi'}}{i}{\cal D}^{(-)}(\pm(s-\tau),\varsigma;\mu_1)=
\nonumber \\
&&\!\!\!\!\!\!\!\!\!\!\!\!\!\!\!\!\!\!
=(\mp 1)\frac{1}{4\pi}\delta_{\xi,\xi'}\left\{\ln\left(i\overline{\mu} 
\frac{\varsigma}{\pi} \sinh \left(\frac{\pi}{\varsigma}
(\pm(s-\tau)-i0)\right)\right) - g\left(\varsigma,\mu_1\right) \right\},
\label{pm_DW} \\
&&\!\!\!\!\!\!\!\!\!\!\!\!\!\!\!\!\!\!
\left[\widetilde{\varphi}^{\xi (\pm)} \left(s; [\pm]\varsigma\right),
\widetilde{\varphi}^{\xi'(\mp)} \left(\tau; [\pm]\varsigma\right)\right]= (\mp 1)
\frac{\delta_{\xi,\xi'}}{i} \widetilde{{\cal D}}^{(-)}(\pm(s-\tau),\varsigma;\mu_1)=
\nonumber \\
&&\!\!\!\!\!\!\!\!\!\!\!\!\!\!\!\!\!\!
= (\mp 1) \frac{1}{4\pi} \delta_{\xi, \xi'} \left\{\ln \left(i \overline{\mu} 
\frac{\varsigma}{\pi} \sinh \left(\frac{\pi}{\varsigma}
(\mp(s-\tau)-i0)\right)\right) - g \left(\varsigma,\mu_1\right) \right\},
\label{pmtt_DW} \\
&&\!\!\!\!\!\!\!\!\!\!\!\!\!\!\!\!\!\!
\left[\varphi^{\xi (\pm)} \left(s; [\pm]\varsigma\right),
\widetilde{\varphi}^{\xi'(\mp)}\left(\tau;[\pm]\varsigma\right)\right] = 
\nonumber  \\
&&\!\!\!\!\!\!\!\!\!\!\!\!\!\!\!\!\!\!
=(\pm 1)[\pm 1] \frac{1}{4\pi}\delta_{\xi, \xi'}
\left\{\ln \left(\cosh \left(\frac{\pi}{\varsigma}(s-\tau)\right)\right)
- f (\varsigma, \mu_2)\right\},
\label{pmt_DW} \\
&&\!\!\!\!\!\!\!\!\!\!\!\!\!\!\!\!\!\!
\left[\varphi^{\xi(\pm)}(s;[\pm]\varsigma),Q^{\xi'(\mp)}([\pm]\varsigma)\right]=
\delta_{\xi, \xi'}\left[\frac{i}{4}-
(\pm 1)\left(\frac{\widehat{x}^0-s}{2\varsigma}\right)\right],
\label{pm_ph_Q} \\
&&\!\!\!\!\!\!\!\!\!\!\!\!\!\!\!\!\!\!
\left[\varphi^{\xi (\pm)} (s; [\pm]\varsigma),
\widetilde{Q}^{\xi' (\mp)} ([\pm]\varsigma)\right] = (\pm 1)[\pm 1]
\delta_{\xi, \xi'} \left(\frac{\widehat{x}^0-s}{2\varsigma}\right),
\label{pm__ph_Qt} \\
&&\!\!\!\!\!\!\!\!\!\!\!\!\!\!\!\!\!\!
\left[Q^{\xi (\pm)} ([\pm]\varsigma), Q^{\xi'(\mp)}([\pm]\varsigma) \right] = 
(\pm 1) a_1 \delta_{\xi, \xi'}
= \left[\widetilde{Q}^{\xi (\pm)}([\pm]\varsigma),
\widetilde{Q}^{\xi'(\mp)}([\pm]\varsigma) \right],
\label{QQ_QtQt} \\
&&\!\!\!\!\!\!\!\!\!\!\!\!\!\!\!\!\!\!
\left[Q^{\xi (\pm)} ([\pm]\varsigma), \widetilde{Q}^{\xi'(\mp)}([\pm]\varsigma) \right]\!
=\! (\pm 1)[\mp 1] a_2 \delta_{\xi, \xi'}\!=
\! \left[\widetilde{Q}^{\xi (\pm)} ([\pm]\varsigma), 
Q^{\xi'(\mp)}([\pm]\varsigma) \right]\!.
\label{QQ_Qt} 
\end{eqnarray}
Here the following quantities are defined: $\overline{\mu}= \mu e^{C_\ni}$, 
\begin{eqnarray}
&&\!\!\!\!\!\!\!\!\!\!\!\!\!\!\!\!\!\!
g\left(\varsigma, \mu_1\right) = \int\limits_{\mu_1}^\infty
\frac{d k^1}{k^0}\left(\frac{2}{e^{\varsigma k^0} - 1}\right) \Longrightarrow
\frac{2}{\varsigma \mu_1} - \ln\left(\frac{2\pi}{\varsigma \overline{\mu}_1}\right),
\quad \overline{\mu}_1= \mu_1 e^{C_\ni} \rightarrow 0, 
\label{ggg0} \\ 
&&\!\!\!\!\!\!\!\!\!\!\!\!\!\!\!\!\!\!
f (\varsigma, \mu_2) = \int\limits_{\mu_2}^\infty \frac{dk^1}{k^0}\,
\frac{1}{\sinh(\varsigma k^0/2)} \Longrightarrow \frac{2}{\varsigma\mu_2}-\ln 2,
\quad \mu_2 \rightarrow 0, 
\label{fff0} \\
&&\!\!\!\!\!\!\!\!\!\!\!\!\!\!\!\!\!\!
\lim_{\varsigma \rightarrow \infty}g \left(\varsigma, \mu_1\right) = 0,\quad 
\lim_{\varsigma \rightarrow \infty}f (\varsigma, \mu_2) = 0,
\label{fff} \\
&&\!\!\!\!\!\!\!\!\!\!\!\!\!\!\!\!\!\!
\delta_L \left(k^1\right)=
\int\limits_{-\infty}^\infty\frac{dx^1}{2\pi}\Delta\left(\frac{x^1}L\right)
e^{\pm ik^1x^1}=L\overline{\Delta} \left(k^1 L\right),\;\;
\lim_{L \rightarrow \infty} \delta_L \left(k^1\right)
= \delta \left(k^1\right), 
\label{DDddDD} \\
&&\!\!\!\!\!\!\!\!\!\!\!\!\!\!\!\!\!\!
a_0 = \pi\!\int\limits_0^\infty d k^1 k^1 \left(\delta_L(k^1)\right)^2\!
= \pi\! \int\limits_0^\infty d t t (\overline{\Delta}(t))^2 \equiv \pi I_1^\Delta,\;\;
I_n^\Delta =\!\int\limits_0^\infty d t t^n\!\left(\overline{\Delta}(t)\right)^2\!,
\label{a_0_I_n} \\
&&\!\!\!\!\!\!\!\!\!\!\!\!\!\!\!\!\!\!
a_1 = a_0+ 2\pi \int\limits_0^\infty d k^1 k^1 
\frac{\left(\delta_L(k^1)\right)^2}{e^{\varsigma k^0}-1}
= a_0 + 2 \pi \int\limits_0^\infty d t t
\frac{(\overline{\Delta}(t))^2}{e^{\varsigma t/L}-1},
\label{a_1_00} \\
&&\!\!\!\!\!\!\!\!\!\!\!\!\!\!\!\!\!\!
a_1\Longrightarrow 2 \pi I_0^\Delta \frac{L}{\varsigma}
+ \frac{\pi}{6} I_2^\Delta \frac{\varsigma}{L}
+ O \left(\left(\frac{\varsigma}{L}\right)^3\right), \quad L\to\infty,\quad 
\lim_{\varsigma \rightarrow \infty} a_1=a_0, 
\label{a_1_0} \\
&&\!\!\!\!\!\!\!\!\!\!\!\!\!\!\!\!\!\!
a_2 = \pi \int\limits_0^\infty d k^1 k^1
\frac{\left(\delta_L(k^1)\right)^2}{\sinh(\varsigma k^0/2)}
= \pi \int\limits_0^\infty d t t
\frac{(\overline{\Delta}(t))^2}{\sinh \left(t\varsigma/2L\right)},
\label{a_2_0} \\
&&\!\!\!\!\!\!\!\!\!\!\!\!\!\!\!\!\!\!
a_2\Longrightarrow 2 \pi I_0^\Delta \frac{L}{\varsigma}
-\frac{\pi}{12} I_2^\Delta \frac{\varsigma}{L} + O
\left(\left(\frac{\varsigma}{L}\right)^3\right),  \quad L\to\infty,
\quad \lim_{\varsigma \rightarrow \infty} a_2=0, 
\label{a_2_}
\end{eqnarray}
where $C_\ni$ is the Euler-Mascheroni constant. It is important to note that in any case 
the difference $a_1-a_2$ becomes $L$ - independent at $L\to\infty$, and if $a_0$ is 
finite, then $a_1-a_2\rightarrow 0$ at $L\to\infty$. 

Following \cite{blot}, by the use of the fields given above, one
can construct a variety of different inequivalent representations of solutions 
of the Dirac equation for a free massless trial field at finite temperature, 
$\partial_\xi\chi_\xi \left(x^{-\xi},\varsigma\right)= 0$ in
the form of local normal ordered exponentials of the left and
right bosonic thermofields $\varphi^\xi(x^\xi,\varsigma)$,
and their charges $Q^\xi(\varsigma)$ (\ref{T_K_7}), (\ref{T_K_7_Q}). 
However, the kinematic independence (\ref{T_vbnmei4}) of the tilde-partners can be 
achieved only by ``admixing'' the Klein factors of both the charges 
$\widetilde{Q}^\xi(\varsigma)$ and $\widetilde{Q}^{-\xi}(\varsigma)$ to the same field. 
Moreover, according 
to the meaning of $L$ as macroscopic parameter, the wanted thermofield should have a 
correct thermodynamic limit $L\to\infty$ for the finite temperature $T>0$. 
The most simple case, which leads to the bosonization relations (\ref{T_nweyi019}), 
(\ref{T_nweyi19}) 
for the currents (\ref{T_bos-111})--(\ref{T_K_Z}) of the fields $\chi(x,\varsigma)$ 
with $Z_{(\chi)}(a)=1$ reads at $L\to\infty$, for $a_1-a_2\rightarrow 0$, 
($\varpi$ and $\Theta$ are arbitrary initial overall and relative phases) as:
\begin{eqnarray}
&&\!\!\!\!\!\!\!\!\!\!\!\!\!\!\!\!\!\!
\chi_\xi (x^{-\xi}; [\pm]\varsigma)={\cal N}_\varphi
\left(\exp\left\{R_\xi (x^{-\xi};[\pm]\varsigma)\right\}\right) 
u_\xi\left(\mu_1,[\pm ]\varsigma\right), 
\label{T_nblaie12} \\
&&\!\!\!\!\!\!\!\!\!\!\!\!\!\!\!\!\!\!
R_\xi (x^{-\xi}; [\pm]\varsigma)= - i2\sqrt{\pi}\left[
\varphi^{-\xi} \left(x^{-\xi}; [\pm] \varsigma\right)
+ \frac{\sigma_0^\xi}{4} {\rm G}^{-\xi}([\pm] \varsigma)
+ \frac{\sigma_1^\xi}{4}  {\rm G}^{\xi}([\pm] \varsigma)\right],
\label{chi_chi} \\
&&\!\!\!\!\!\!\!\!\!\!\!\!\!\!\!\!\!\!
u_\xi\left(\mu_1,[\pm]\varsigma\right)=
\left(\frac{\overline{\mu}}{2\pi}\right)^{1/2}e^{i\varpi - i\xi\Theta/4}
\exp\left\{-\,\frac{g(\varsigma,\mu_1)}{2}\right\},
\label{T_nblaie13} \\
&&\!\!\!\!\!\!\!\!\!\!\!\!\!\!\!\!\!\!
\mbox{with: }\;
\sigma^\xi_0=- \xi\sigma, \quad \sigma^\xi_1=\xi 1+\rho,
\label{sig_s_r}
\end{eqnarray}
where the $\sigma$ and $\rho$ are defined by the condition (\ref{T_vbnmei4}), 
and the new charges, with simple commutation relations following from 
(\ref{pm_ph_Q})--(\ref{QQ_Qt}), are used: 
\begin{eqnarray}
&&\!\!\!\!\!\!\!\!\!\!\!\!\!\!\!\!\!\!
{\rm G}^{\xi}([\pm]\varsigma)=Q^{\xi}([\pm]\varsigma)+[\pm 1]
\widetilde{Q}^{\xi}([\pm]\varsigma), \mbox { with:}
\label{G_Gt} \\
&&\!\!\!\!\!\!\!\!\!\!\!\!\!\!\!\!\!\!
\left[\varphi^{\xi(\pm)}(s;[\pm]\varsigma),{\rm G}^{\xi'(\mp)}([\pm]\varsigma)\right]=
\frac{i}{4}\,\delta_{\xi, \xi'}, 
\label{cc_phi_G_G} \\
&&\!\!\!\!\!\!\!\!\!\!\!\!\!\!\!\!\!\!
\left[\varphi^{\xi(\pm)}(s;[\pm]\varsigma),
\widetilde{\rm G}^{\xi'(\mp)}([\pm]\varsigma)\right]=
[\pm 1]\frac{i}{4}\,\delta_{\xi,\xi'}, 
\label{cc_phi_Gt} \\
&&\!\!\!\!\!\!\!\!\!\!\!\!\!\!\!\!\!\!
\left[{\rm G}^{\xi(\pm)}([\pm]\varsigma),{\rm G}^{\xi'(\mp)}([\pm]\varsigma)\right]=
(\pm 1)2(a_1-a_2)\delta_{\xi,\xi'}, 
\label{cc_phi_G_Gt} \\
&&\!\!\!\!\!\!\!\!\!\!\!\!\!\!\!\!\!\!
\left[{\rm G}^{\xi(\pm)}([\pm]\varsigma),
\widetilde{\rm G}^{\xi'(\mp)}([\pm]\varsigma)\right]=
(\pm 1)[\pm 1]2(a_1-a_2)\delta_{\xi,\xi'}. 
\label{cc_G_Gt}
\end{eqnarray}
Following to \cite{ks_t}, we obtain then at $L\to\infty$ the normal exponential of the DM
for Thirring field in the form analogous to \cite{blot,oksak} ($\Lambda$ is ultraviolet 
cut-off):
\begin{eqnarray}
&&\!\!\!\!\!\!\!\!\!\!\!\!\!\!\!\!\!\!
\Psi_\xi (x; [\pm]\varsigma)={\cal N}_\varphi\left(\exp\left\{
\Re_\xi(x;[\pm]\varsigma)\right\}\right) w_\xi\left(\mu_1,\varsigma\right), 
\label{T_nweyi29} \\
&&\!\!\!\!\!\!\!\!\!\!\!\!\!\!\!\!\!\!
\Re_\xi(x;[\pm]\varsigma)=-i\left[\Xi^{-\xi}(x;[\pm]\varsigma)+ 
\frac{\Sigma_0^\xi}{4}{\rm G}^{-\xi}([\pm]\varsigma)+
\frac{\Sigma_1^\xi}{4}{\rm G}^{\xi}([\pm]\varsigma)\right], 
\label{T_Psi_Psi} \\
&&\!\!\!\!\!\!\!\!\!\!\!\!\!\!\!\!\!\!
\Xi^{-\xi}(x;[\pm]\varsigma)=
\overline{\alpha}\varphi^{-\xi}\left(x^{-\xi};[\pm]\varsigma\right)+
\overline{\beta}\varphi^{\xi}\left(x^\xi;[\pm]\varsigma\right), 
\label{T_Psi_vrho} \\
&&\!\!\!\!\!\!\!\!\!\!\!\!\!\!\!\!\!\!\!\!
w_\xi(\mu_1,\varsigma)=\left(\frac{\overline{\mu}}{2\pi}\right)^{1/2}
\left(\frac{\overline{\mu}}{\Lambda}\right)^{\overline{\beta}^2/4\pi }
e^{i\varpi- i \xi\Theta/4}
\exp\left\{-g(\varsigma,\mu_1)
\left(\frac{1}{2}+\frac{\overline{\beta}^2}{4\pi}\right)\right\},
\label{T_nweyi30} \\
&&\!\!\!\!\!\!\!\!\!\!\!\!\!\!\!\!\!\!
\mbox{where: }\;
\Sigma^\xi_0=\overline{\alpha}\sigma^\xi_0-\overline{\beta}\sigma^\xi_1, \quad 
\Sigma^\xi_1=\overline{\alpha}\sigma^\xi_1-\overline{\beta}\sigma^\xi_0. 
\label{Sig_Sig} \\
&&\!\!\!\!\!\!\!\!\!\!\!\!\!\!\!\!\!\!
\mbox{with: }\;\sigma^\xi_0 \Rightarrow \xi(2n_0+1), \quad 
\sigma^\xi_1 \Rightarrow \xi 1+(2n_2+1),\quad n_0,\,n_2\,\mbox{ -- integer},
\label{sgm_n_n} 
\end{eqnarray}
by imposing again the conditions onto the parameters that are necessary to have correct 
Lorentz-transformation properties corresponding to the spin $1/2$, and correct canonical 
anticommutation relations (\ref{ZLa_Z0}) respectively: 
\begin{eqnarray}
\overline{\alpha}^2-\overline{\beta}^2= 4\pi, \quad
\overline{\beta} - \frac{\beta g}{2\pi} = 0. 
\label{T_nweyi32}
\end{eqnarray}
Remarkably, that the obtained conditions (\ref{sgm_n_n}) provide the anticommutation 
(\ref{ZLa_Z0}), locality and kinematic independence relations (\ref{T_vbnmei4}) for both 
the free (\ref{T_nblaie12}) and Thirring fields (\ref{T_nweyi29}) and their tilde 
partners simultaneously. 

Straightforward calculation of the vector current operators 
(\ref{T_bos-111})--(\ref{T_K_Z}) for the solution (\ref{T_nweyi29})--(\ref{sgm_n_n}), 
with $Z_{(\Psi)}(a)=(-\Lambda^2 a^2)^{-{\overline{\beta}^2}/{4\pi}}$, $Z_{(\chi)}(a)=1$, 
by means of Eqs. (\ref{T_K_9})--(\ref{QQ_Qt}) and (\ref{T_nweyi32}), under the 
conditions \cite{ks_t}:
\begin{eqnarray}
&&\!\!\!\!\!\!\!\!\!\!\!\!\!\!\!\!\!\!
\overline{\alpha}=\left(\frac{2\pi}{\beta}+
\frac{\beta}{2}\right),\quad
\overline{\beta}=\left(\frac{2\pi}{\beta}-
\frac{\beta}{2}\right),\;\mbox{ with: }\;
\frac{2\sqrt{\pi}}{\beta}=\sqrt{1+\frac{g}{\pi}},
\label{T_Kab}
\end{eqnarray}
again directly reproduces the bosonization (and linearization) relations 
(\ref{T_ns94m61}), (\ref{T_ns94m62}), (\ref{T_nweyi019}), (\ref{T_nweyi19}) as the 
following weak equalities:
\begin{eqnarray}
&&\!\!\!\!\!\!\!\!\!\!\!\!\!\!\!\!\!\!
\widehat{J}_{(\Psi)}^\nu(x,\varsigma)\stackrel{\rm w}{=}
-\,\frac{\beta}{2\pi}\,\epsilon^{\mu\nu}\partial_\nu\phi(x, \varsigma)=
\frac{\beta}{2\sqrt{\pi}}\widehat{J}_{(\chi)}^\nu(x,\varsigma),
\label{T_K_Za} 
\end{eqnarray}
demonstrating self-consistency of the above calculations. The obtained normal form of 
Thirring thermofields has a correct renormalization properties:
\begin{eqnarray}
&&\!\!\!\!\!\!\!\!\!\!\!\!\!\!\!\!\!\!\!\!
\left\{\Psi_\xi(x,\varsigma),\Psi_{\xi'}^\dagger(y,\varsigma)\right\}\Bigr|_{x^0= y^0}=
Z_{(\Psi)}(x-y)\Bigr|_{x^0= y^0}\delta_{\xi,\xi'}\delta\left(x^1-y^1\right),
\label{ZLa_Z0} \\
&&\!\!\!\!\!\!\!\!\!\!\!\!\!\!\!\!\!\!\!\!
\mbox{with: }\; Z_{(\Psi)}(x-y)\Bigr|_{x^0= y^0}=
\left[\Lambda^2(x^1-y^1)^2\right]^{-{\overline{\beta}^2}/{4\pi}}\sim 1, 
\label{ZLa_Z}
\end{eqnarray}
for $x^1-y^1\simeq 1/\Lambda$. The limit of this solution to zero temperature generalizes 
Oksak solution \cite{blot,oksak} as two-parametric Thirring field with arbitrary 
continuous parameters $\sigma$ and $\rho$ from (\ref{sig_s_r}). 

\section{Conclusion} 

The main lesson of our work is very simple: the correct true HF should be only a fully 
normal ordered operator in the sense of DM onto irreducible physical fields. 
Only this form clarifies and assures 
correct renormalization, commutation and symmetry properties. It allows also a simple 
connections between different types of solutions with finite and zero temperature. 
The chosen here representation space of free massless pseudoscalar field relax the 
problem of nonpositivity of inner product. 
Contrary to the recent works \cite{abr,abr3}, we take into account different types of 
charge regularization and all possible mutual commutation relations of bosonic 
thermofields and their charges, that self-consistently removes fictitious 
$\widehat{x}^0$ -dependence. 

The authors thank Y. Frishman, A.N. Vall, S.V. Lovtsov, V.M. Leviant, 
and participants of seminar in LTPh JINR for useful discussions.

This work was supported in part by the RFBR (project N 
09-02-00749) and by the program ``Development of Scientific 
Potential in Higher Schools'' (project N 2.2.1.1/1483, 
2.1.1/1539).


\begin{thebibliography}{00}
\bibitem {isih} 
  {\it Isihara A.}, Statistical physics (Academic Press, New York -- London, 1971).
\bibitem {gom_ste}
  {\it Gomez Nic\'ola A., Steer D.,} Thermal bosonization in the sine-Gordon and Massive 
  Thirring Models. // Nucl.Phys. {\bf B549}, 1999, 409--449.
\bibitem {perelom}
  {\it Perelomov A.M.}, Generalized coherent states and their applications 
  (Springer, Berlin, 1986).
\bibitem {lipkin}
  {\it  Lipkin H.J.}, Quantum mechanics (North-Holland Publishing Company, Amsterdam, 
  1973).
\bibitem {blot}
  {\it  Bogoliubov N.N., Logunov A.A.,  Oksak A.I.,  Todorov I. T.},
  General principles of quantum field theory (Kluwer Academic Publishers, Boston, 1990).
\bibitem{oksak}
  {\it  Oksak A.I.}, Non-Fock linear boson systems and their applications in
  two-dimentional models. //  Teoret. Mat. Fiz. {\bf 48}, 1981, 297--318.
\bibitem{d_f_z}
  {\it Dell'Antonio G.F., Frishman Y., Zwanziger D.}, Thirring model in terms of
  currents: solution and light-cone expansions. // Phys. Rev. {\bf D6}, 1972, 988--1007.
\bibitem{ks_t}
    {\it  Korenblit S.E.,  Semenov V.V.},  Massless Thirring model in canonical 
    quantization scheme. // J. Nonlin. Math. Phys., {\bf 18}, N 1, 2011, 65-74.  
    {\it  Korenblit S.E.,  Semenov V.V.}, Integration of the Thirring model equations. // 
    Russian Physics Journal, {\bf 53}, N 6, 2010, 630--638. 
    (arXiv: hep-th/1003.1439 v.2)
\bibitem{mtu}
  {\it  Umezawa H., Matsumoto H., Tachiki M.,} Thermo-field dynamics and condensed states
  (North-Holland Publishing Company, Amsterdam, 1982).
\bibitem{ojima}
  {\it Ojima I.}, Gauge fields at finite temperatures: thermo field dynamics,
  KMS condition and their extension to gauge theories. // Annals of Physics {\bf 137}, 
  1981, 1--32.
\bibitem{alv-gom}
  {\it Alvarez-Estrada R.F., Gomez Nic\'ola A.}, The Schwinger and Thirring models at 
  finite chemical potential and temperature. // Phys. Rev. {\bf D57}, 1998, 3618--3633.
\bibitem{abr}
  {\it Amaral R.L.P.G., Belvedere L.V., Rothe K.D.}, Two-dimensional
  thermofield bosonization. // Annals of Physics, {\bf 320}, 2005, 399--428.
\bibitem{abr3}
   {\it Amaral R.L.P.G., Belvedere L.V., Rothe K.D.,  Rodrigues A.F.}, 
  Quantum electrodynamics in two dimensions at finite  temperature: thermofield 
  bosonization approach. // J.Phys. {\bf A44}, 2011, 025401.
\end{thebibliography}
\end{document}